\documentclass[fleqn]{2023SCGE}
\usepackage{amsmath}
\usepackage{booktabs}
\usepackage{array}
\usepackage{footnote}
\makesavenoteenv{tabular}
\usepackage{listings}
\usepackage{multirow}

\begin{document}

\ensubject{subject}
\ArticleType{Article}
\Year{2024}
\Month{March}
\Vol{67}
\No{3}
\DOI{10.1007/s11433-023-2270-7}
\ArtNo{230412}
\ReceiveDate{August 17, 2023}
\AcceptDate{November 3, 2023}
\OnlineDate{January 29, 2024}

\title{Advancing space-Based gravitational wave astronomy: Rapid parameter estimation via normalizing flows}{Advancing space-Based gravitational wave astronomy: Rapid parameter estimation via normalizing flows}


\author[1]{Minghui Du$^\dagger$}{}
\author[2,3,5]{Bo Liang$^\dagger$}{}
\author[4,5]{He Wang}{hewang@ucas.ac.cn}
\author[1, 2, 5, 6]{Peng Xu}{}
\author[1, 2, 5]{Ziren Luo}{}
\author[4,7,8]{Yueliang Wu}{ylwu@itp.ac.cn}

\AuthorMark{M. Du}

\AuthorCitation{M. Du, B. Liang, H. Wang, P. Xu, Z. Luo, and Y. Wu}

\address[1]{Center for Gravitational Wave Experiment, National Microgravity Laboratory, Institute of Mechanics, Chinese Academy of Sciences, Beijing 100190, China}
\address[2]{Key Laboratory of Gravitational Wave Precision Measurement of Zhejiang Province, Hangzhou Institute for Advanced Study, \\ University of Chinese Academy of Sciences, Hangzhou 310024, China}
\address[3]{Shanghai Institute of Optics and Fine Mechanics, Chinese Academy of Sciences, Shanghai 201800, China}
\address[4]{International Centre for Theoretical Physics Asia-Pacific, University of Chinese Academy of Sciences, Beijing 100049, China}
\address[5]{Taiji Laboratory for Gravitational Wave Universe (Beijing/Hangzhou), University of Chinese Academy of Sciences, Beijing 100049, China}
\address[6]{Lanzhou Center of Theoretical Physics, Lanzhou University, Lanzhou 730000, China}
\address[7]{Hangzhou Institute for Advanced Study, University of Chinese Academy of Sciences, Hangzhou 310024, China}
\address[8]{Institute of Theoretical Physics, Chinese Academy of Sciences, Beijing 100190, China}
\contributions{These authors contributed equally to the work.}

\abstract{  
Gravitational wave (GW) astronomy is witnessing a transformative shift from terrestrial to space-based detection, with missions like Taiji at the forefront. 
While the transition brings unprecedented opportunities for exploring massive black hole binaries (MBHBs), it also imposes complex challenges in data analysis, particularly in parameter estimation amidst confusion noise. 
Addressing this gap, we utilize scalable normalizing flow models to achieve rapid and accurate inference within the Taiji environment. 
Innovatively, our approach simplifies the data’s complexity, employs a transformation mapping to overcome the year-period time-dependent response function, and unveils additional multimodality in the arrival time parameter.
Our method estimates MBHBs several orders of magnitude faster than conventional techniques, maintaining high accuracy even in complex backgrounds. 
These findings significantly enhance the efficiency of GW data analysis, paving the way for rapid detection and alerting systems and enriching our ability to explore the universe through space-based GW observation.
}



\maketitle

\begin{multicols}{2}

\section{Introduction}\label{Sec:intro}
The burgeoning field of gravitational wave (GW) astronomy has been significantly enhanced through current advancements in terrestrial gravitational wave detection, led by initiatives such as LIGO–Virgo–KAGRA network~\cite{aasi2015advanced,acernese2014advanced,kagra2019kagra,abbott2020prospects}.
\Authorfootnote
\noindent 
Ground-based observatories have uncovered a new 
dimension of understanding the universe, revealing the ripples in spacetime produced by violent astrophysical processes.
However, the future of gravitational wave research is set to transcend terrestrial constraints, with ambitious space-based detection missions on the horizon. 
Pioneering programs, such as LISA~\cite{amaro2017laser,baker2019laser}, Taiji~\cite{hu2017taiji,luo2021taiji,2021CmPhy...4...34T_simplified}, and TianQin~\cite{luo2016tianqin}, represent the vanguard of this next scientific frontier.

The Taiji project, a space-based gravitational wave detection initiative in China, was started in 2008 and has matured into a full-fledged mission over years of preliminary research.
Endorsed by the Chinese Academy of Sciences, it is projected to make its inaugural launch in the 2030s~\cite{ziren2020taiji}.
{\color{black}Taiji consists of a triangle of three spacecraft (S/Cs) with a baseline separation of 3 million kilometers. The constellation orbits around the Sun and leads the Earth by about $20^\circ$. 
Taiji is proposed to detect low-frequency GWs emitted by sources such as compact galactic binaries (GBs), massive black hole binaries (MBHBs), extreme mass ratio inspirals (EMRI), as well as the stochastic gravitational wave background of astrophysical or cosmological origins, covering the frequency range of 0.1 mHz to 1.0 Hz. } Renowned for its diverse scientific objectives, Taiji holds the potential to open up new avenues in the fields of astronomy, cosmology, and fundamental physics~\cite{TPFp,DM}. Its exploration may ultimately lead to breakthroughs in understanding the elusive nature of quantum gravity~\cite{QG}.

At present, with the anticipated approval of the Taiji-2 mission and the continuous enhancement of Taiji's sensitivity~\cite{luo2020brief,luo2022recent}, it is projected that an increasing number of gravitational wave events will be discovered, paving the way for a deeper understanding of our universe.

As we transition towards the era of space-based gravitational wave detection, new challenges in data processing begin to surface.
A significant challenge in this domain lies in the non-trivial overlaps of individual signals in both the time and frequency domains~\cite{bayle2022overview,speri2022roadmap}.
This problem is compounded by factors such as the diversity of long-lived sources in the targeted millihertz frequency band, including GBs, MBHBs, and EMRIs. 
The sensitive frequency band of Taiji will likely be swamped with $\sim 10^4$ resolvable sources, each contributing to a high-dimensional parameter space that must be constrained. This complexity not only presents a tremendous data analysis challenge~\cite{PhysRevD.72.022001}, but also further exacerbates the problem by introducing uncertainties in the types and numbers of sources. The inability to predetermine the templates to be used in the traditional matched filtering method prior to the analysis adds to the complexity. These factors combine to create a daunting task that demands innovative solutions, as not only is reaching an unbiased solution highly challenging, but the computational cost of tackling this multifaceted problem is also predictably expensive, as evidenced in numerous studies~\cite{cornish2020black,karnesis2023eryn,Weaving:2023fji,strub2023accelerating}.

Current solutions to this problem, termed `global fit' techniques~\cite{cornish2005lisa,littenberg2020global,littenberg2023prototype}, aim to perform a joint fit to all sources in the data and noise, an algorithmically and computationally challenging task.
While dedicated pipelines will be developed for different source classes, the role of a global-fit pipeline becomes crucial.
This pipeline will serve as an interface between different source pipelines, orchestrating joint updates to all source and noise parameters to prevent biases and source confusion.
A prototype of such a pipeline is presented in the work of Littenberg and Cornish~\cite{littenberg2023prototype}, where they conduct a global fit analysis of the LISA Data Challenge (LDC) Sangria dataset~\cite{Baghi:2022ucj}.

However, we cannot overlook the computational demands of existing parameter estimation techniques of individuals.
The prototype global fit pipeline~\cite{littenberg2023prototype} currently requires as many as $\mathcal{O}(10^3)$ CPUs and several days to process the 12-month LDC data, and the issue of computational resources would become even more critical for the multi-year datasets in the future.
Thus, it is apparent that the development of a more efficient framework for quickly inferring gravitational waves is a pressing need in the field.

While existing global-fit techniques can handle the dense overlapping of signals that characterize space-based GW data, individual pipelines are still indispensable, especially for identifying unique events. 
The mergers of MBHBs provide a case in point, as they often display signal-to-noise ratios (SNR) of $10^2$ to $10^3$ and appear as distinct peaks in the time series of data~\cite{pratten2023lisa}. These signals, although challenging to discern, can still be individually resolved with targeted processing.
In this context, the role of individual source pipelines becomes crucial.
These pipelines serve as a pre-processing step, delving deeper into the data with a focus on specific source types.
They operate on the most recent best-fit residual from the global fit and provide feedback on any identified sources, which are then included in future refinements of the global solution.
This iterative process ensures a continuously updated and more comprehensive global fit as additional data is received from the satellites.


In light of the challenges and advancements discussed, and with an emphasis on the quest for efficiency and speed, this paper builds upon our recent research that utilized deep learning methods for rapid detection~\cite{RUAN2023137904} and parameter estimation~\cite{ruan2023parameter} of MBHB sources. 
In our recent parameter estimation work~\cite{ruan2023parameter}, we focused on the partial inference of MBHB sources without confusion noise and explored the challenges and opportunities in learning the reference time over a whole year.
The current work serves as a summary and significant advancement of our past experiences.
Three core highlights of this paper include:

\begin{enumerate}
    \item Achieving a complete and unbiased 11-dimensional fast statistical inference in the presence of confusion noise, a complex background that has posed challenges for space-based gravitational wave observation as compared to terrestrial detection.
    \item Addressing the unique challenge posed by Taiji in comparison to ground-based detectors, where the year-period time-dependent response function enhances the complexity of the data. This complexity poses difficulties for neural network training. To tackle this, we employ prior knowledge of the symmetry in Taiji's response function to propose a transformation mapping between different reference times. This strategy enables us to train a model on a simplified dataset while conducting parameter estimation on data gathered throughout the whole mission period, effectively navigating the complex Taiji environment.
    \item Discovery that the Normalizing Flow (NF) method, when inferring the posterior distribution, brings an additional multimodality to the arrival time parameter in the extrinsic parameters.
    A detailed examination of this phenomenon could be highly beneficial and meaningful for future alerting systems, enhancing our ability to quickly respond to significant events.
\end{enumerate}

By diverging from traditional methods and employing a scalable NF-based model, we apply this innovative approach within the Taiji environment for parameter estimation of MBHBs. 
Our method achieves results in several orders of magnitude faster than traditional methods while maintaining high accuracy and unbiasedness. 
Importantly, our approach demonstrates robust performance, even in the presence of confusion noise due to unresolvable galactic binaries.
Thus, this work contributes a valuable preprocessing step for the global fit, significantly enhancing the efficiency of gravitational wave data analysis and providing new avenues for rapid detection and multi-dimensional parameter inference.

Except for acting as a preprocessing stage of global fit, a fast detection and inference technique also allows for the discovery of potential electromagnetic (EM) counterparts that occur in the post-merger phase of MBHBs, such as the disc rebrightening, the formation of an X-ray corona, and that of an incipient jet~\cite{LISA:2022yao,Gold:2014dta}.
To alert and guide the search for such EM counterparts, it is crucial that reliable estimates for the sky location and distance can be done with low latency. 


This paper is organized as follows: Section \ref{Sec:model} reviews the applications of deep learning on gravitational wave astronomy, as well as an in-depth discussion of the method developed in this study.
Section \ref{Sec:data} presents a detailed account of the prior setup, data generation process, and other relevant configurations used in our work.
Section \ref{Sec:result} offers a comprehensive presentation of the results derived from our methodology, while section \ref{Sec:final} rounds out the paper with a conclusion and offers perspectives on future developments in this area. 
Following this logic, the paper systematically unfolds the innovative approach and findings that contribute to the advancement of space-based gravitational wave detection.

\section{Methodological framework}\label{Sec:model}

In this section, we begin with a review of the methodology adopted in this study, particularly the use of machine learning techniques in fast and accurate gravitational-wave inference.

Machine learning techniques~\cite{jordan2015machine} 
has been widely applied within the gravitational wave community, leading to remarkable results~\cite{cuoco2020enhancing}.
Specifically, deep learning~\cite{2015LeCunDeeplearninga}
has advanced significantly in the inference application for ground-based gravitational wave data processing~\cite{gabbard2022bayesian,chatterjee2019using,green2020gravitational,green2021complete,delaunoy2020lightning,krastev2021detection,shen2021statistically,dax2021real,dax2023neural}, notably more so than in space-based gravitational wave parameter estimation~\cite{chua2020learning,ruan2023parameter}.
This progress can be attributed primarily to the technique of normalizing flow~\cite{kobyzev2020normalizing,papamakarios2021normalizing}, which is used for rapid inference.

The accelerated inference capabilities of normalizing flow have opened up numerous research applications.
For instance, its rapid processing advantage has been leveraged to examine the detection performance of overlapping signals in third-generation ground-based gravitational wave detectors~\cite{langendorff2023normalizing}.
In the arena of population inference~\cite{ruhe2022normalizing}, NF has been utilized for high-dimensional parameter estimation.
Further, it's been ingeniously combined with traditional methods~\cite{williams2021nested} to maximize the strengths of both approaches, paving the way for more nuanced analyses.
Lastly, NF has been directly employed in tests of general relativity~\cite{crisostomi2023neural}, providing a fresh perspective and potentially unveiling new insights.
These diverse applications demonstrate the versatile potential of NF technology, solidifying its integral role in the continued advancement of gravitational wave detection and data processing.

In the following subsections, we will delve into the specifics of how we have leveraged this cutting-edge technology in our research.

\subsection{Introduction to the normalizing flow}

Normalizing flow~\cite{kobyzev2020normalizing,papamakarios2021normalizing} is a framework for generative models that transforms a simple base distribution into a more complex posterior distribution through a series of invertible transformations.
These invertible transformations are typically implemented using neural networks and ensure both the invertibility of the transformations and the ability to compute the Jacobian matrix.
During the inference process, normalizing flow generates samples by applying a sequence of invertible transformations to the base distribution.
The objective of inference is to learn the parameters of these transformations, enabling the mapping of the base distribution to a posterior distribution that closely matches the true data distribution.


In recent years, there have been numerous applications of normalizing flows in the scientific domain. Conor Durkan et al.~\cite{durkan2019neural} introduced Neural Spline Flows, a methodology that employs piecewise spline transformations to model complex data probability distributions, enabling parameter inference tasks.
Inspired by Neural Spline Flows, Stephen R Green et al.~\cite{green2021complete} applied the approach to ground-based gravitational wave parameter inference, demonstrating remarkable results~\cite{dax2021real,dax2023neural}.
We have drawn inspiration from the above-mentioned works and developed an innovative and scalable composite normalization flow model for gravitational wave parameter inference.

\subsection{Overview on the neural spline flow }

Neural spline flow~\cite{durkan2019neural} is a generative model that is based on normalizing flows.
It uses a neural network architecture to model the probability density function of a given dataset.
The key idea behind neural spline flow is to approximate the invertible transformation between a simple base distribution (usually a multivariate Gaussian) and the target distribution using piecewise rational quadratic functions called splines.
These splines are used to model the non-linear mapping between the input and output spaces, allowing for more flexible and expressive modeling.

The key mathematical concept behind piecewise rational quadratic functions is the use of rational quadratic polynomials.
A rational quadratic polynomial is a ratio of two quadratic polynomials. The forward transformation has the form:
\begin{equation}
y(x) = \frac{ax^2+bx+c}{dx^2+ex+f}\,,
\end{equation}
where $a,b,c,d,e,f$ are parameters that control the shape of the function.
The piecewise nature of this function arises from dividing the input space into different regions.
Each region is associated with a different set of parameters, allowing the function to capture different patterns in the data.

In the context of neural spline flow, these piecewise rational quadratic functions are used as invertible transformations between a simple distribution and a more complex one.
The parameters of the quadratic segments are learned through training neural networks, enabling the model to capture complex dependencies in the data.

\subsection{Customization for the Taiji scenario: A scalable approach}

In our quest for a more comprehensive analysis of Taiji's gravitational wave data, we need to account for the detector noise Power Spectral Density (PSD), denoted as $S_n(f)$, or equivalently the Amplitude Spectral Density (ASD), defined as $A_n(f) \equiv \sqrt{S_n(f)}$.  They play indispensable roles in data generation and greatly impact the result of data analysis.
As such, we incorporated PSD as additional contextual information within our neural network, represented as $q(\theta|d, S_n)$, where $d$ represents the input data containing the MBHB signal and other noise, and $\theta$ represents the parameters to be inferred for the MBHB.
By considering previously estimated PSDs, we can adapt the network during the inference process, thus enabling standard ``off-source" noise estimation.

\begin{figure}[H]
  \centering
  \includegraphics[width=0.5\textwidth]{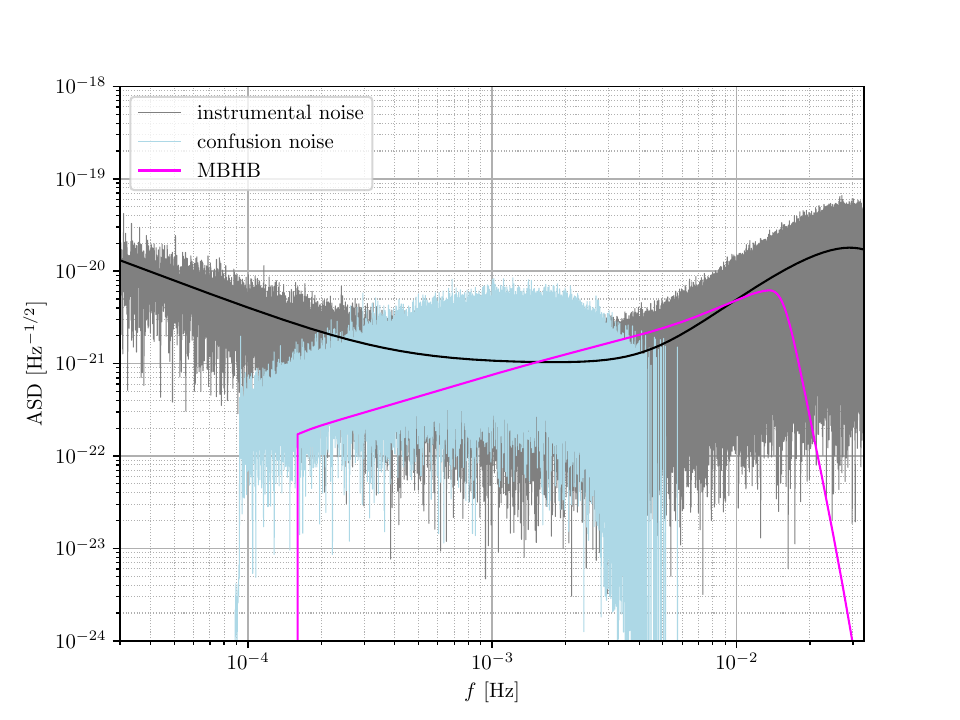}
  \caption{
  ASD plot of representative Taiji data spanning a full 12-month duration, exhibited in the Time Delay Interferometry (TDI) A channel.
  The instrument noise is depicted in gray, generated from the Taiji noise model.
  The light blue curve manifests the foreground noise arising from numerous Galactic binaries (GBs).
  The magenta line displays a case of Massive Black Hole Binary (MBHB) inspiral-merger-ringdown waveform drawn from our prior.
  This spectrum provides a representative overview of the varied sources of noise and signals in our dataset, underlining the complexity of the analytical scenario.
  }
  \label{fig:TestSample}
\end{figure}

A notable challenge we encountered during this process was handling high-dimensional observed data.
In our experiments, we analyzed data characterized by three dimensions: (2, 3, 2877).
These dimensions are specific to our study and encompass 2 Time Delay Interferometry (TDI) variables $\{A, E\}$, 3 data channels that include the real and imaginary parts of the Fourier-domain data along with the real-value ASD, and finally, 2877 distinct frequencies (see Figure~\ref{fig:TestSample}).
Clearly, the multiplication of these dimensions results in an incredibly high dimensionality. 
To address this issue, we incorporated an embedding neural network model specifically designed to extract feature vectors from such complex data sets.
This scalable neural network model brings a significant benefit: it can adapt and adjust the composition of residual blocks~\cite{he2016deep} based on the specific characteristics of the dataset and PSD data for various parameter inference tasks.
A comprehensive summary of the key hyperparameters used in our model configuration can be found in Table~\ref{table:hyperparameters} in~\ref{appendix:modelconfig}.

For each event, the time when the merger waveform reaches the Solar System Barycenter (SSB) is denoted as $t_c$.
The constellation of Taiji orbits the sun with the period of a year, hence resulting in a year-period time-dependent response function. 
For the completeness of the training dataset, the range of $t_c$ should span at least one year. 
Such complexity of data presents a great challenge for the training of the neural network,  which would subsequently impact the performance of parameter estimation.

In the face of this problem, our earlier work~\cite{ruan2023parameter} took a step back and was confined to estimating only 4 parameters.
While, in this paper, to directly address this issue, we exploit our prior knowledge of the symmetry of Taiji's response function,  
{\color{black} reasonably simplify the data, }
and expand the {\color{black}application} of deep learning parameter estimation to the whole 11-dimensional parameter space.
To be specific, the neural network is trained on a dataset generated within a small range of $t_c$ centered around a reference time $t_{\rm ref}$.
Then, at the inference stage, after a few manipulations of the data and posterior samples  (see Section~\ref{subsec:generalization} for the details), the trained neural network can be generalized to make inferences on data near another reference time, say $t_{\rm ref}^\prime$.  Our solution to the complexity of data integrates the symmetry of the system into the ``black box" created by deep learning,  showing a promising prospect of the methods belonging to this family, such as~\cite{Dax:2021myb}.

This transformation, based on certain assumptions, allows us to infer the posterior distribution of parameters at any other reference time, which is a significant advantage in practice.
The integration of these adaptations and improvements has been instrumental in advancing our application of neural network models for the Taiji scenario.

\section{Data manipulation and experiment setup}\label{Sec:data}

In this section, we delve into the intricate process of data preparation and experiment setup, laying the foundation for our normalizing flow application.
We first detail the generation of waveform signals, detector noise model, TDI settings, and preprocessing tasks required for our training data.
Following this, we discuss the practical aspects of our training implementation, focusing on the data utilization, sampling, and optimization of hyperparameters.



\subsection{Data generation and preprocessing}

In this section, we describe our recipe to simulate the MBHB datasets used for model training and validation. 
Throughout this paper, we work in the frequency domain, enabling a fast calculation of the detector responses to incident GWs. 
As the necessity of using TDI in suppressing laser frequency noise has been widely acknowledged, our data is generated in the form of noise orthogonal TDI variables ${A, E}$, which are derived from the original 1st generation Michelson channels ${X, Y, Z}$. 
Therefore, the modeling of Taiji's responses involves two steps: the simulation of waveforms and their conversion into the output of $A, E$ channels. 
The gravitational wave response of the Taiji detector. These steps can be schematically summarized as 
\begin{align}
    A, E(f) = \sum_{\alpha} \mathcal{T}^{A, E}_{\alpha}(f) \tilde{h}_{\alpha}(f), \quad \alpha \in \{+, \times\},
\end{align}
where $\mathcal{T}^{A, E}_{\alpha}(f)$ is often referred to as the transfer function.

For the first step, the IMRPhenomD model is employed to describe the dominant harmonics of the waveforms emitted by MBHBs with aligned spins. 
As for the second step, we apply the fast Fourier domain response provided by Ref.~\cite{bbhx_Katz:2020hku,bbhx_Katz:2021uax,bbhx_London2018,bbhx_Marsat:2018oam,bbhx_Marsat:2020rtl,bbhx_michael}, which has a close dependence on the orbits of the detector.  In order to generate a large amount ($\sim 10^6$) of samples at a fast computational speed while retaining essential features of the data, the motion of Taiji's satellites is modeled based on an equal-arm analytic orbit. In this simplified orbit, the SSB-frame positions of the S/Cs (labeled by $i \in {1 ,2, 3}$) read
\color{black}\begin{align}
    x_i(t) &= R\cos\alpha + \frac{L}{2\sqrt{3}}\left[\frac{1}{2}\sin2\alpha\sin\gamma_i  
     - (1+\sin^2\alpha)\cos\gamma_i\right], \nonumber \\
    y_i(t) &= R\sin\alpha + \frac{L}{2\sqrt{3}}\left[\frac{1}{2}\sin\alpha\cos\gamma_i 
     - (1+\cos^2\alpha)\sin\gamma_i\right], \nonumber \\
    z_i(t) &= -\frac{L}{2}\cos\left[\alpha - \gamma_i\right], \nonumber \\
    \alpha(t) &= \frac{2\pi }{T}t + \alpha_0, \nonumber \\
    \gamma_i &= \frac{2\pi (i  - 1)}{3} + \gamma_0,
\end{align}
where $L=3\times 10^9 {\rm m}$ is the nominal arm length of Taiji, $R, T$ equal 1 Astronomical Unit and 1 year, respectively, and the initial conditions are set as $\alpha_0 = \gamma_0 = 0$. 

Overall, the TDI response to each MBHB signal can be fully specified by 11 parameters: the chirp mass $\mathcal{M}_c$, mass ratio $q \equiv m_2 / m_1$, the dimensionless spins along the $z-$axis $\chi_{z1}$ and $\chi_{z2}$, the SSB time $t_c$ and phase $\varphi_c$ at coalescence, the luminosity distance $d_L$, the inclination angle $\iota$, polarization angle $\psi$, and the ecliptic coordinate $\{\lambda, \beta\}$, being the longitude and latitude respectively.

\begin{table*}[t]
    \footnotesize
    \begin{threeparttable}
    \tabcolsep 10pt 
    \caption{Prior Distributions used in this work.
    The table provides the ranges of prior distributions for each parameter with their lower and upper bounds.
    For each parameter, a uniform distribution is assumed between these bounds.
    The lower (upper) bound for the luminosity distance is based on a redshift of 1 (10.0).
    For inclination ($\iota$) and ecliptic latitude ($\beta$), we consider distributions that represent isotropic angles on the sphere.}\label{table:priors}
    \begin{tabular*}{\textwidth}{ccccc}
    \toprule
        \hline
        \hline
        \textbf{Parameter} & \textbf{Description} & \textbf{Prior Lower Bound} & \textbf{Prior Upper Bound} & \textbf{Alias in Code}\\
        \hline
        $\mathcal{M}_c$ & Chirp mass of the binary system & $2.5 \times 10^5 M_\odot$ & $1.25 \times 10^6 M_\odot$ & \lstinline|chirp_mass|\\
        $q$ & Ratio of masses of the two black holes & 0.125 & 1.0 & \lstinline|mass_ratio|\\
        $\chi_{z1}$ & Spin of the 1st black hole along the $z$-axis & 0 & 0.99 & \lstinline|spin_1|\\
        $\chi_{z2}$ & Spin of the 2nd black hole along the $z$-axis & 0 & 0.99 & \lstinline|spin_2|\\
        $t_c$ & Time of coalescence relative to the reference time\tnote{1)}, $t_{\rm ref}$ & -0.01 day & 0.01 day & \lstinline|coalescence_time|\\
        $\varphi_c$ & Phase at the moment of coalescence & 0 rad & $2\pi$ rad & \lstinline|coalescence_phase|\\
        $d_L$ & Luminosity distance to the binary system & 6000 Mpc & 100000 Mpc & \lstinline|luminosity_distance|\\
        $\iota$ & Angle of inclination of the binary orbit & 0 rad & $\pi$ rad & \lstinline|inclination|\\
        $\beta$ & Ecliptic latitude of the binary system & $-\frac{\pi}{2}$ rad & $\frac{\pi}{2}$ rad & \lstinline|latitude|\\
        $\lambda$ & Ecliptic longitude of the binary system & 0 rad & $2\pi$ rad & \lstinline|longitude|\\
        $\psi$ & Polarization angle of the gravitational wave & 0 rad & $\pi$ rad & \lstinline|psi|\\
        \hline
        \hline
        \bottomrule
    \end{tabular*}
        \begin{tablenotes}
        \item[1)] The reference time $t_{\rm ref}$ is set to the $30^{\rm th}$ day.
        \end{tablenotes}
    \end{threeparttable}
\end{table*}

The priors on these parameters are listed in {\color{black}Table~\ref{table:priors}}.
The two mass parameters $\mathcal{M}_c$ and $q$ are derived from the uniform distributions of component masses $m_i \ (i = 1, 2)$.
Similar to the LDC-sangria data, the prior range of $\mathcal{M}_c$ covers approximately an order of magnitude. 
As a demonstration of the new method, currently the model is only trained on positive spin values. In order to conduct a more realistic study, and considering that IMRPhenomD is only an incomplete description of the MBHB waveform, which does not take into account the subdominant harmonics and unaligned spins, our future work should be based on the whole parameter space, and adopt waveforms such as IMRPhenomXHM~\cite{Garcia-Quiros:2020qpx} or IMRPhenomPXHM~\cite{Pratten:2020ceb}. This means that the complexity of  data would increase significantly, which might raise be necessity of improving or refining the model, while the discussion on this aspect is beyond the scope of this paper. 
Parameter $t_c$ is defined relative to the reference time $t_{\rm ref}$, which is set to the $30^{\rm th}$ day for the training dataset, without loss of generality. 
Since the mergers of MBHBs manifest as prominent peaks in the time series of data, we consider the possibility of constraining the merger time within a prior range of 0.02 days after some primary detecting procedures that can roughly determine the times of mergers.
Besides, the range of $d_L$ corresponds to the redshift range of $z \in [1, 10]$, which is sufficient to encompass the majority of the  MBHBs observable by space-based GW detectors.
The angles $\iota$ and $\beta$ are uniformly distributed in terms of $\cos\iota$ and $\sin \beta$ to represent that the sources are isotropically distributed on the sphere.

\begin{figure}[H] 
  \centering
  \includegraphics[width=0.5\textwidth]{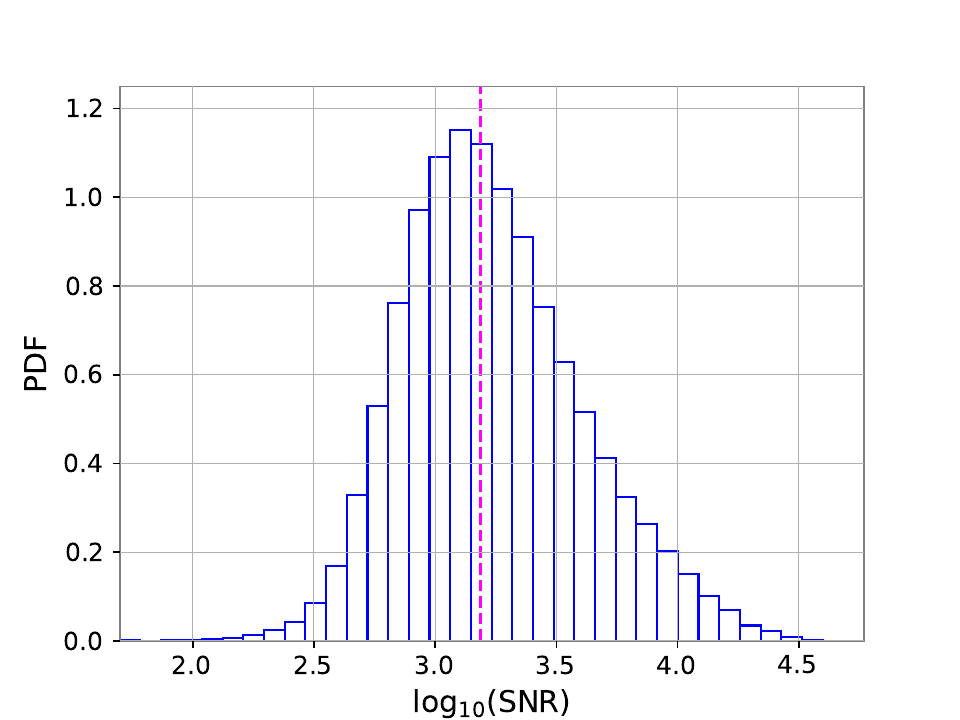}
  \caption{The distribution of SNRs in the training set. The SNR of our test event is shown as a magenta dashed line .}
  \label{fig:SNRPDF}
\end{figure}

Shown in Figure~\ref{fig:SNRPDF} is the distribution of SNRs in the training set. The magnitudes of SNRs span a wide range from $10^2$ to $10^4$, which is the typical range of Lisa/Taiji's targeted MBHBs. The test event used in Sec.~\ref{Sec:result} (the magenta line in Figure~\ref{fig:TestSample}) is marked with a magenta dashed line near the peak of distribution, indicating that it is a representative event for the dataset.

MBHBs are known to emit GW signals that reach their peak signal-to-noise ratio over short durations~\cite{pratten2023lisa}, typically within a single day during the merger phase. 
To reasonably avoid the potential overlap of different merger signals, the time window for each sample is set to one day, with the merger positioned towards the end of this window.
The aforementioned feature of MBHB waveforms ensures that enough information can be retained within a one-day time window.
Furthermore, the sampling frequency is set as $1/15$ Hz, indicating a Nyquist frequency of $0.033$ Hz.
This frequency exceeds the highest instantaneous frequency of the waveforms, ensuring comprehensive capture of the signal.

Along with the responses, we also need the PSD of the noise in each channel.
For instrumental noises, according to the current design of the Taiji mission, at the targeted frequency around $1 \ {\rm mHz}$, the displacement noise and residual acceleration noise are expected to be $8\ {\rm pm / \sqrt{{\rm Hz}}}$ and $3\ {\rm fm/s^2/\sqrt{{\rm Hz}}}$, respectively.
The one-sided amplitude special density of noises is proposed in a unit of the fractional frequency 
 as~\cite{luo2020brief}
\begin{equation}\label{eq:TaijiNoise}
    \begin{aligned}
        \sqrt{S_{\mathrm{OMS}}(f)}= & 8 \times 10^{-12} \frac{2 \pi f}{c} \sqrt{1+\left(\frac{2 \mathrm{mHz}}{f}\right)^4} \frac{1}{\sqrt{\mathrm{Hz}}}, \\
        \sqrt{S_{\mathrm{ACC}}(f)}= & 3 \times 10^{-15} \frac{1}{2 \pi f c} \\
        & \times \sqrt{1+\left(\frac{0.4 \mathrm{mHz}}{f}\right)^2} \sqrt{1+\left(\frac{f}{8 \mathrm{mHz}}\right)^4} \frac{1}{\sqrt{\mathrm{Hz}}},
    \end{aligned}
\end{equation}
where $S_{\rm OMS}(f)$ and $S_{\rm ACC}(f)$ stand for the noise originating from the optical measurement system and acceleration of test masses.
 The total instrumental noise PSDs of the $A, E$ channels $S_n^{A, E}(f)$ are the combinations of these two components (see Ref.~\cite{Vallisneri:2007xa,Wang:2020vkg,Wang:2020fwa} for their specific forms).
For simplicity, we treat $S_n(f)$ as constant over the observation duration. 
Our model is trained on datasets composed of ``clean" TDI responses and stationary instrumental noise. On a more realistic ground, the long-term variations and short-term jitters of instrumental noise, along with the typical anomalies such as glitches and gaps should be taken into consideration in future work.

On the other hand, to test the robustness and generalizability of the model, during the inference stage, we do take into consideration a non-stationary noise caused by the unresolvable foreground originating from the overlapping GWs of GBs.
This confusion noise is generated from a catalog of $\sim 3\times10^7$ GBs provided by the LDC.  We calculate the TDI responses of these enormous sources with the open-source code GBGPU~\cite{GBGPU_Cornish:2007if,GBGPU_Robson:2018svj,GBGPU_michael_l_katz_2022_6500434}, which is modified to fit for the orbit and arm-lengths of Taiji. Among the whole catalogue, $\sim 2\times 10^4$ sources with SNR greater than 7 were removed, to simulate an ideal situation where these bright sources have already been successfully identified and isolated. The residual unresolvable GBs then add up to form the confusion noise.
Our choice of the SNR threshold follows the convention of~\cite{Wang:2023jct,2019MNRAS.483.5518K,liu2023confusion,Zhang:2022wcp}.

\subsection{Training implementation and optimization}

Prior to the actual training phase, we generated and stored a set of 5 million waveforms on a hard drive. 
This preparatory step served to create a robust base for subsequent computational tasks.
In the training phase itself, the parameters of the luminosity distance and coalescence time were continuously sampled.
This strategy was chosen to ensure generalizability without loss of essential detail.

The training involved a total of 600 epochs, with a batch size of 10240.
We started with an initial learning rate of 0.00005 and employed a combination of cosine annealing~\cite{loshchilov2016sgdr} and the Adam optimizer~\cite{kingma2014adam} for the gradual reduction of the learning rate to zero over the course of the training process.
To avoid overfitting, 5\% of the dataset was reserved for validation, and throughout this process, no indications of overfitting were detected.
The training was executed using a single NVIDIA A800 GPU and was completed in approximately 6 days.

\section{Results and discussion}\label{Sec:result}
\subsection{Generalization of the network}\label{subsec:generalization}

\begin{figure}[H] 
  \centering
  \includegraphics[width=0.5\textwidth]{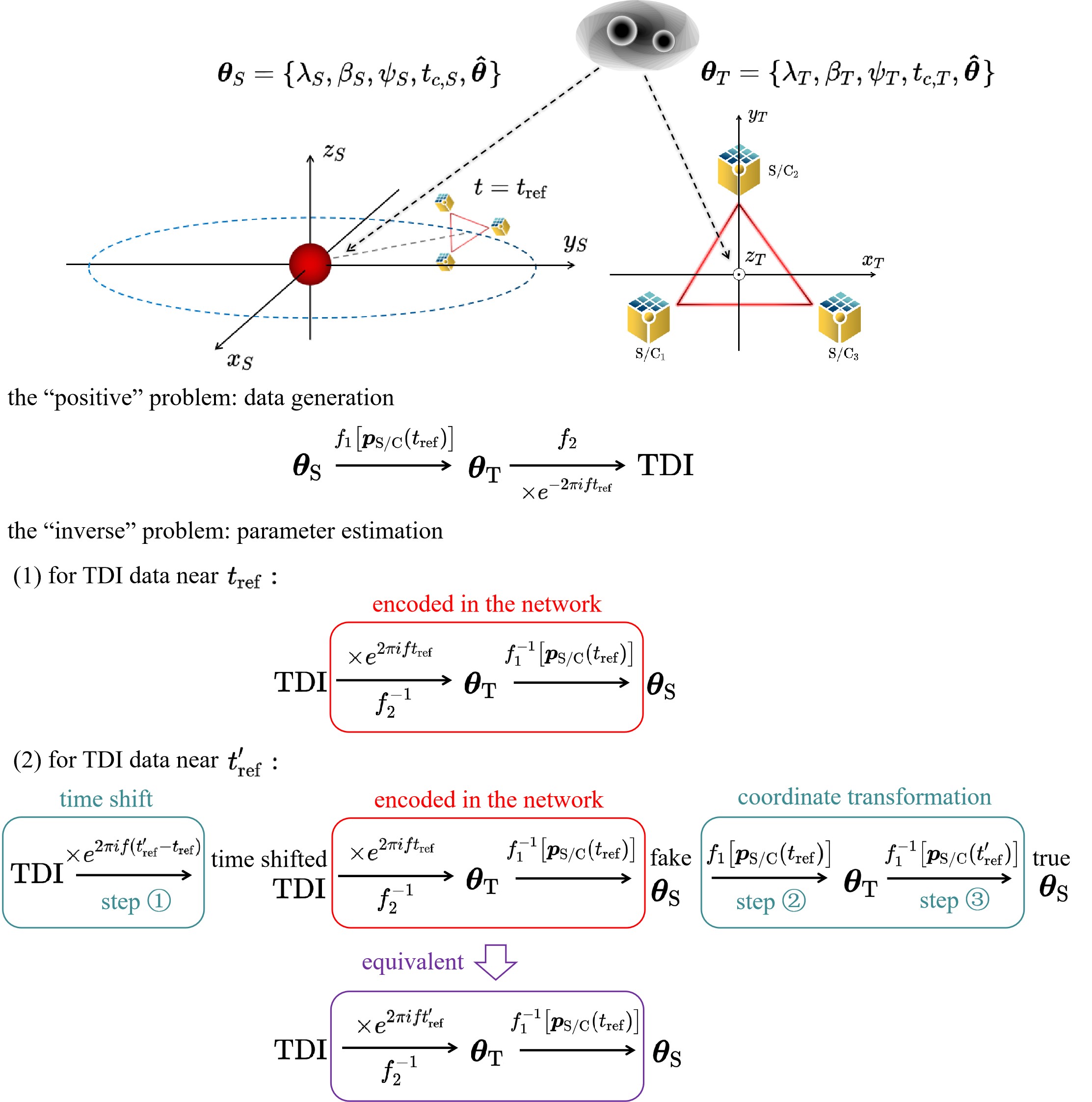}
  \caption{
A schematic illustration depicting the generalization of our network during the inference stage.
The top section of the illustration shows the SSB and Taiji frames, with two black dashed arrows symbolizing not two separate GW signals, but rather indicating how the sky location and arrival time of the same GW signal take different values in these two frames. Here, $\boldsymbol{\hat{\theta}}$ represents the parameters that remain invariant under a coordinate transformation.
For a concise and clear understanding, we exclude probabilistic terms and approach parameter estimation as the inverse problem of data generation. This ``positive" problem translates the SSB-frame parameters to their Taiji-frame counterparts via a time-dependent mapping $f_1$, then to the TDI outputs through a time-independent mapping $f_2$, and an exponential term. The procedure of parameter estimation thus unravels these mappings in reverse.
Upon training on the dataset, the network encodes the inference rule at time $t_{\rm ref}$, as depicted in the red box. To perform parameter estimation at a different reference time $t_{\rm ref}^\prime$, both the data and posterior samples undergo manipulation according to the three steps outlined in Section~\ref{subsec:generalization}, demonstrated in the blue boxes. Specifically, the initial two exponential terms amalgamate into a single expression $e^{2\pi i f t_{\rm ref}^\prime}$, and the mappings $f_1$ and $f_1^{-1}$ at $t_{\rm ref}$ nullify each other. As a result, we essentially reach the purple box, which embodies the needed rule for inference at $t_{\rm ref}^\prime$.
}
  \label{fig:generalization}
\end{figure}

At the beginning of this section, we explain the method of generalizing our network. Figure~\ref{fig:generalization} gives a more intuitive demonstration of the whole scheme.
As previously noted, to simplify the training task, our dataset is generated within a moderately narrow range of $t_c$, centered around $t_{\rm ref}$, and the trained model can be generalized to make inferences on data near another reference time $t_{\rm ref}^\prime$. 
To illustrate this, we define two coordinate frames, i.e., the SSB frame, as explained before, and the Taiji frame, whose origin is placed at the center of the constellation. 
Parameters that differ in these two frames are $\{t_c, \lambda, \beta, \psi\}$, where $t_c$ should be interpreted as the time when the merger waveform reaches the origin of the frame relative to some $t_{\rm ref}$. 
We distinguish their values in these 2 frames by subscripts ``S" (for SSB) or ``T" (for Taiji). 
Other parameters, denoted by $\boldsymbol{\hat{\theta}}$, stay the same in both of the frames (see Figure~\ref{fig:generalization}).

An important concept is that, in the Taiji frame, for short data durations (e.g. one day), the relative movement of GW sources due to the motion and rotation of Taiji can be safely neglected. For example, Ref.~\cite{Cornish:2021smq} demonstrated that, in the context of a first-stage analysis, LISA-like detectors can be treated as static in one-month data segments. Therefore, the mapping from $\boldsymbol{\theta}_T$ to the TDI waveform can be divided to a time-independent part (dubbed $f_2$), and a factor $e^{-2\pi i f t_{\rm ref}}$, accounting for the effect of $t_{\rm ref}$.

Moreover, given the positions $\boldsymbol{p}_{\rm S/C}(t_{\rm ref})$ of S/Cs at some $t_{\rm ref}$, the transformation (dubbed $f_1[\boldsymbol{p}_{\rm S/C}(t_{\rm ref})]$) between $\boldsymbol{\theta}_S$ and $\boldsymbol{\theta}_T$ can be done analytically (See Ref.~\cite{Vallisneri:2012np} for the specific formalism). 
Consequently, even if the network has only learned the time-dependent relationship between $\boldsymbol{\theta}_S$ and the TDI response at a specific $t_{\rm ref}$ (the $30^{\rm th}$ day in our case), with the aid of coordinate transformation, it has essentially learned the time-invariant mapping $f_2$, and can be then generalized to make parameter estimation at any other reference time.

It is worth noting that our method relies on analytical orbits, and while using numerically simulated orbits would not impede the coordinate transformation itself, it could theoretically pose questions about real-world applicability or efficiency. However, in practice, these concerns are alleviated as the coordinate transformation, involving only simple matrix operations, is relatively efficient. For example, transforming 50,000 samples takes slightly over a minute, but this time can be significantly reduced using parallel processing with CPU or GPU. Therefore, the method maintains its suitability and effectiveness for actual implementation without significant efficiency concerns.

To implement this concept at the inference stage, we manipulate the data and posterior samples following 3 steps.  
Firstly, the data is time-shifted from $t_{\rm ref}^\prime$ to $t_{\rm ref}$ by multiplying a factor of $e^{2\pi i f (t_{\rm ref}^\prime - t_{\rm ref})}$,
and we make inferences on this shifted data with the trained network to get a set of ``fake" samples. 
The term ``fake" indicates that the parameter estimation is based on the transfer function at $t_{\rm ref}$, rather than $t_{\rm ref}^\prime$.
Secondly, we convert the fake parameters into their Taiji-frame values based on the coordinate transformation at $t_{\rm ref}$. Finally, the parameters are converted back to the SSB frame, but this time with the coordinate transformation at $t_{\rm ref}^\prime$. Now the samples can be regarded as representing the  ``true" posterior.  Unless otherwise specified, the results shown in this section are all based on the above operations. 

The generalization process illustrated in this section emphasizes the adaptability of our network to different reference times.
It also underscores the capability of the network to handle complex transformations that occur between different coordinate frames. 
This thorough examination of the network's generalization properties enriches our understanding of its performance and robustness in the context of gravitational wave analysis.

\subsection{Unbiased estimation and confidence validation}

The primary objective of this part of our work is not merely validation but an in-depth examination of our algorithm's ability to ensure unbiased estimation of the parameters, reinforcing the robustness of our methods.
We achieve this by conducting a Kolmogorov-Smirnov (KS) test~\cite{veitch2015parameter} to compare one-dimensional distributions output by our algorithms. 

\begin{figure}[H]
  \centering
  \includegraphics[width=0.45\textwidth]{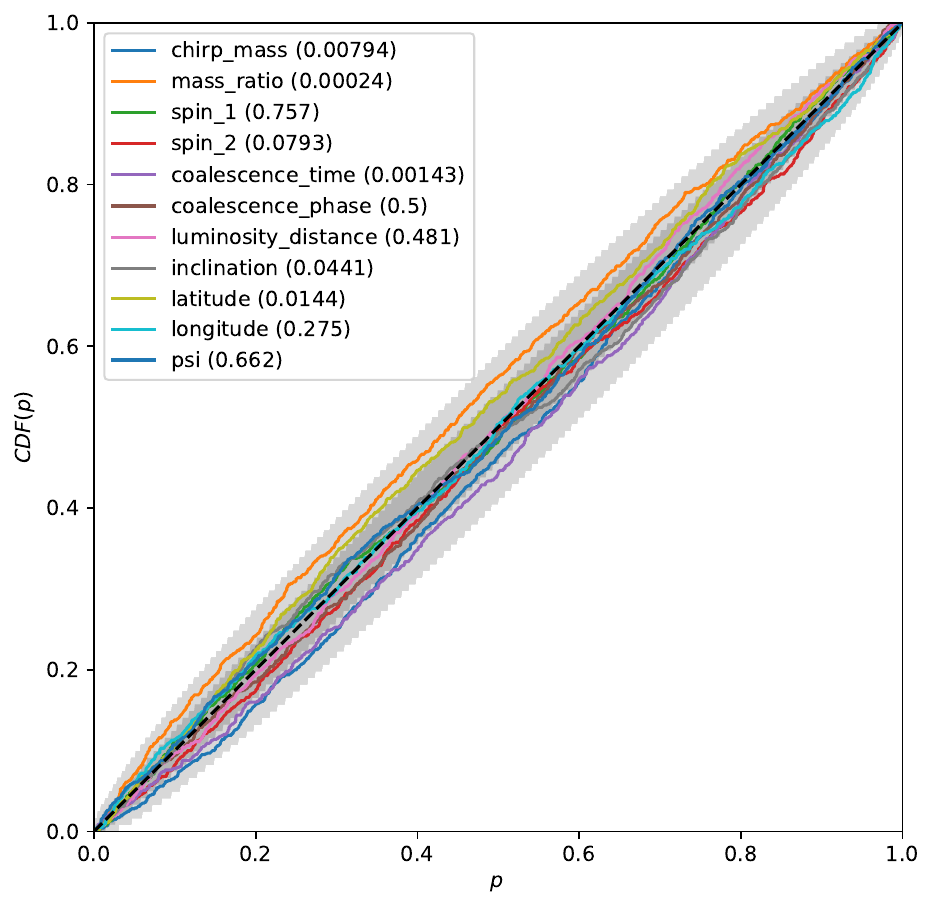}
  \caption{
    P-P plot for a set of 1000 injections with added confusion noise and reference times between from 1 to 365 days, as analyzed by our neural network model.
    The plot showcases the cumulative distribution function (CDF) of the percentile scores of true values for each parameter from the marginalized one-dimensional posterior distribution.
    Each percentile value of the injected parameter is computed for each injection and one-dimensional posterior distribution.
    Ideally, the CDF of the injections for each parameter should lie close to the diagonal, indicating proper network performance, as the percentiles should uniformly distribute between 0 and 1.
    Deviations from the diagonal signify discrepancies in the model's ability to learn the true posterior for that parameter.
    Here, the closeness of the CDF to the diagonal confirms the algorithm's accurate sampling of the posteriors and supports the unbiased estimation principle of our method.
    The grey regions demarcate the $1\sigma$ and $2\sigma$ confidence bounds.
    KS test p-values are denoted in the legend.
  }
  \label{fig:p-p plot}
\end{figure}

To further corroborate the validity of our algorithm and assert that the probability distributions we recover genuinely reflect the confidence we should hold in the signal parameters, we undertook an extensive set of tests on simulated signals.
By injecting 1000 waveforms drawn from the prior, along with added confusion noise~\cite{liu2023confusion} and a reference time varying between 1 and 365 days, we gauged the frequency at which true parameters resided within a certain confidence level.
This assessment not only provides empirical evidence of our credible intervals being well-calibrated but also exemplifies the thoughtful examination of our model towards unbiased estimation, ensuring a seamless connection between theory and practice, as shown in Figure~\ref{fig:p-p plot}.

\subsection{Multimodality in extrinsic parameters}

The detailed examination of intrinsic parameters such as chirp mass, mass ratio, and spins, along with extrinsic parameters that are beneficial for alerting, can be revealed through the use of corner plots.

We have the ability to immediately illustrate the 90\% confidence interval for both our machine learning predictions and the masses and spins recovered through parameter inference.
This can be achieved by directly plotting the two-dimensional and one-dimensional marginalized posteriors generated using output samples from our neural network model and stochastic sampling approaches, superimposed on each other.
Since the NF aims to mimic the Bayesian posterior distributions of parameters, its performance can be better illustrated if the result of Bayesian inference on the same event is also presented as a benchmark. 
In this context, we have employed Nested Sampling~\cite{10.1214/06-BA127,PhysRevD.81.062003,Ashton_2019} as a Monte Carlo technique for comparative experimentation. 
This specific choice enables a robust comparison with our NF model, providing insights into the differences and similarities between traditional stochastic sampling methods and our innovative machine-learning techniques.

An example of the illustration, using the same prior in Table~\ref{table:priors} for a test injection with confusion noise on the 30th reference day, is depicted in Figure~\ref{fig:corner}. 
The injected GW signal is emitted by a merger MBHB with total SNR = 1543, whose parameters are shown in Table~\ref{table:injection}.
For the given case, the intrinsic parameters were recovered within the 90\% confidence interval of the injected values, demonstrating the effectiveness of our method. 
Other extrinsic parameters like the sky positions were well recovered, and the correct sky mode was identified.
However, an exception was found in the extrinsic parameter $\varphi_c$, which did not fall within the 90\% confidence interval (as detailed in~\ref{appendix:full_corner}).
This notable discrepancy in $\varphi_c$ can be explained as the network's inability to capture the intrinsic multimodal behavior of this parameter. Indeed, since the IMRPhenomD waveform only includes the dominant (2, 2) mode, the multimodality of $\varphi_c$ is intrinsic to this waveform, while capturing this complex feature seems too challenging for our current network.
Such nuances emphasize the complexity of parameter estimation in gravitational wave analysis and point to areas for further investigation and refinement.


As with the extrinsic parameters, a prominent feature shown in the corner plots of NF is the multimodality of $t_c$, $\beta$, and $\lambda$, a phenomenon that is unseen in the results of nested sampling.
Most notably, the extra peak of $t_c$ is rarely observed in similar works using the stochastic sampling method~\cite{Marsat:2020rtl, Weaving:2023fji}.
By referring to the results of nested sampling, we do not expect significant multimodality in the ecliptic coordinates.

\begin{figure}[H]
  \centering
  \includegraphics[width=0.5\textwidth]{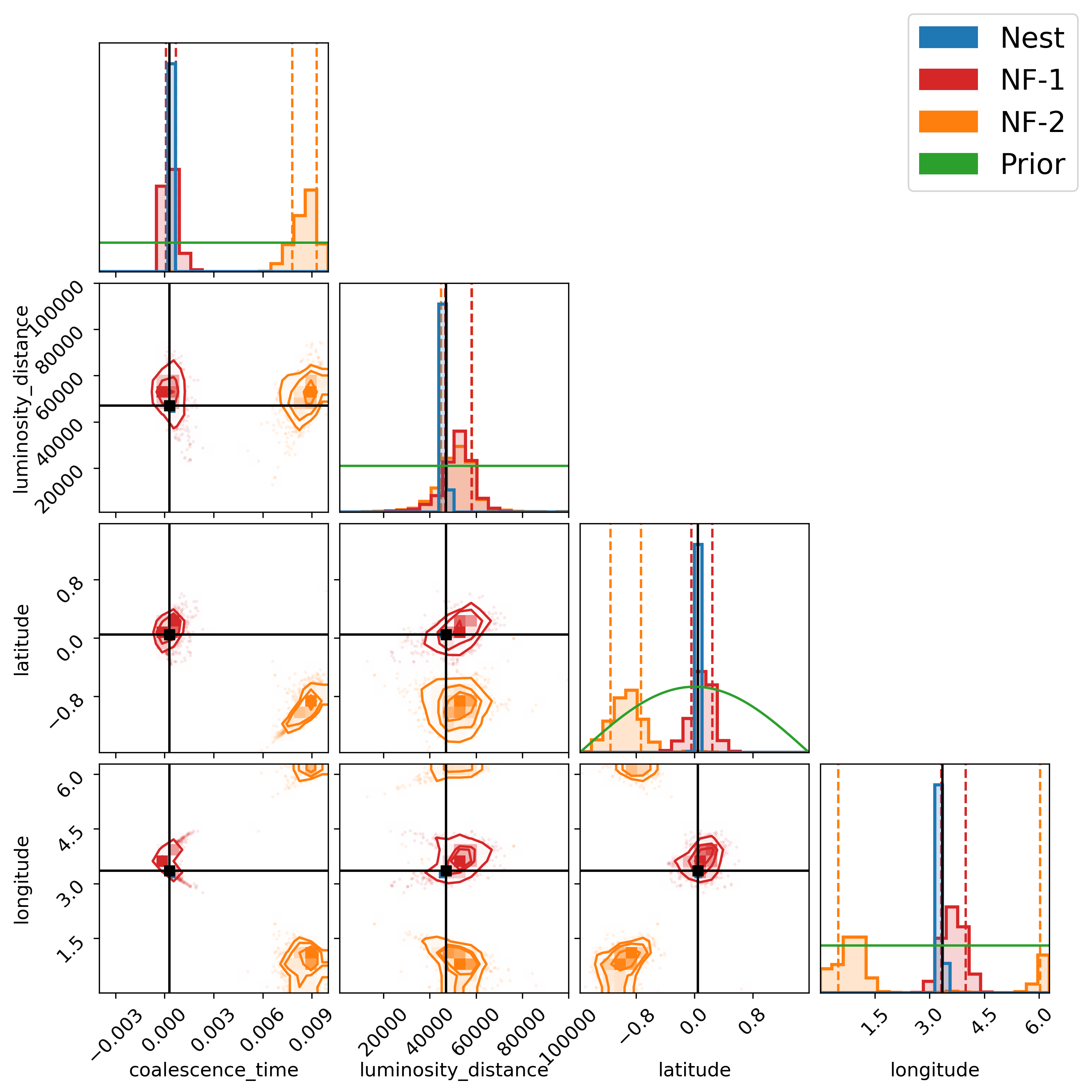}
  \caption{
    Extrinsic parameter posterior corner plot of injected signal in confusion noise with GBs.
    The red and orange contours illustrate the two-dimensional joint posteriors acquired from our normalizing flow (NF) model, with the peaks of $t_c$ labeled as NF-1 (dominant or best-fit) and NF-2 (subdominant), whereas the blue contour denotes the corresponding posteriors yielded from our benchmark analyses using Nested Sampling (Nest), a specific Monte Carlo technique.
    For each method, the contour boundaries encapsulate the $1\sigma$ level.
    One-dimensional histograms of the posterior distribution for each parameter from both methods are plotted along the diagonal, with the prior distribution also displayed in Green.
    The true parameter values of the simulated signal are indicated by black vertical and horizontal lines.
  }
  \label{fig:corner_extrinsic}
\end{figure}

In Figure~\ref{fig:corner_extrinsic}, we distinguish between the two peaks in $t_c$, marking them as NF-1 (dominant or best-fit) and NF-2 (subdominant). This distinction allows for a clearer understanding of how this unique multimodality in NF affects the extrinsic parameters' posterior.
A quite possible explanation, which is discernible from Figure~\ref{fig:corner_extrinsic}, is that the NF may not be powerful enough to provide a conclusive inference on $t_c$, giving rise to a lower peak away from the true value.

This multimodality is subsequently passed on to $\varphi_c$, since they both appear in the phase term of the waveform, and then to $\psi$, $\beta$, and $\lambda$ due to inherent degeneracies.
However, it is important to emphasize that despite these imperfections in the NF posterior, the best-fit values are close to the true values within the $1\sigma$ range for most of the parameters (and at least $2\sigma$ for others).
The existence of other peaks can often be safely ignored, and future work might involve refining the model to mitigate the impact of these subdominant peaks.
Potential solutions may include more targeted training of the NF on specific regions of the parameter space or employing additional constraints to suppress the unwanted multimodality. 
These steps could lead to even more accurate and reliable parameter estimation.

We also have collated the injection's true values and the results yielded by both the model and the nested sampling process in Table~\ref{table:injection}.
This not only provides a concise overview of the performance of each method but also offers valuable insights into how well they are capable of approximating the actual values, such as the constraint on the reference time $t_c$ or other extrinsic parameters.
One notable observation is that our machine learning pipeline generally produces broader posteriors compared to the Bayesian approach.
The discrepancy is approximately an order of magnitude, but when compared to the prior distributions, the results are evidently adequate, signifying the robust performance of our machine-learning model.

This comparison not only validates the efficacy of our approach in estimating intrinsic parameters but also highlights how these constrained posterior results, such as the limitations on $t_c$ time or other extrinsic parameters, can be instrumental.
These constraints are particularly useful for global-fit processing, aiding in preliminary assessments or alerts for gravitational wave detection. 
Furthermore, they indicate areas where further refinement may be needed for some of the extrinsic parameters.
The insights derived from these results are clear, and concise, and will play a vital role in advancing techniques for the detection and analysis of gravitational waves.

\begin{table}[H]
    \footnotesize
    \begin{threeparttable}
    \doublerulesep 0.1pt \tabcolsep 7pt 
    \caption{A comparison between the injected parameters from the prior and the parameters recovered by both the Nested Sampling (Nest) and Normalizing Flow (NF) methods (the first peak, NF-1).
    The recovered values are accompanied by their 90\% confidence regions.
    Although there is approximately an order of magnitude difference between the two methods, the distributions essentially overlap, indicating a close match.
    Despite this difference, both methods provide results that are remarkably accurate when compared to the prior distribution, underscoring their effectiveness.
    }\label{table:injection}
    \begin{tabular}{cccc}
    \toprule
        \textbf{Parameter} & \textbf{Injected value} & \textbf{NF-1} & \textbf{Nest}\\
        \hline
        $\mathcal{M}_c$ [$M_\odot$]   & 531067.4  & $528117.8_{-3665.8}^{+3732.9}$ & $531035.6_{-90.8}^{+102.4}$ \\
        $q$                         & 0.6355 & $0.6492_{-0.0294}^{+0.0291}$ & $0.6412_{-0.0059}^{+0.0053}$ \\
        $\chi_{z1}$ & 0.2930 & $0.1962_{-0.1499}^{+0.1436}$ & $0.2619_{-0.0238}^{+0.0269}$ \\
        $\chi_{z2}$ & 0.2726 & $0.3626_{-0.2231}^{+0.2114}$ & $0.3233_{-0.0424}^{+0.0380}$ \\
        $t_c$ [day]\tnote{1)} & 0.00028 & $0.0003_{-0.0002}^{+0.0004}$ & $0.0003_{-0.0000}^{+0.0000}$ \\
        $\varphi_c$ [rad] & 0.5412 & $2.9889_{-1.9167}^{+2.0764}$ & $3.6876_{-3.1135}^{+1.5906}$ \\
        $d_L$ [Mpc] & 47054.1 & $53161.0_{-6126.3}^{+5002.6}$ & $46277.6_{-672.8}^{+687.2}$ \\
        $\iota$ [rad] & 0.8207 & $0.7324_{-0.1510}^{+0.0737}$ & $0.8303_{-0.0080}^{+0.0075}$ \\
        $\beta$ [rad] & 0.0459 & $0.1264_{-0.1672}^{+0.1232}$ & $0.0395_{-0.0155}^{+0.0143}$ \\
        $\lambda$ [rad] & 3.3544 & $3.6955_{-0.3436}^{+0.2893}$ & $3.3339_{-0.0124}^{+0.0148}$ \\
        $\psi$ [rad] & 1.5634 & $1.5649_{-1.4423}^{+0.1591}$ & $1.5708_{-1.5638}^{+0.0146}$ \\
        \bottomrule
    \end{tabular}
        \begin{tablenotes}
        \item[1)] The reference time $t_{\rm ref}$ is set to the $30^{\rm th}$ day.
        \end{tablenotes}
    \end{threeparttable}
\end{table}

\begin{table*}[t]
\footnotesize
    \begin{threeparttable}
    \tabcolsep 6pt
    \caption{Comparison of computational time required by different posterior sampling approaches to generate their respective samples.}\label{table:efficiency}
        \begin{tabular*}{\textwidth}{ccccc}
        \toprule
        \hline
        \hline Sampling Method & Number of Posterior Samples & Runtime (seconds) & Time per Sample (seconds)\tnote{1)} & Time Ratio `Nest/NF' \\
        \hline Nested Sampling (Nest) & 11,215 & 3,000 & 0.26750 & 0.991$\times10^{3}$ \\
        Normalizing Flow (NF) & 10,000 & 2.7 & 0.00027 & 1 \\
        \hline
        \hline
        \bottomrule
        \end{tabular*}
    \begin{tablenotes}
    \item[1)] The column `Time per Sample' denotes the average computational time taken to generate a single sample.
    \end{tablenotes}
    \end{threeparttable}
\end{table*}

\subsection{Computational performance}

In the final part of our results, we delve into the computational efficiency of our algorithm.
The most computationally intensive aspect of utilizing Normalizing Flows is undoubtedly the training time, which typically takes around one week to conclude.
However, it is crucial to note that once the network is adequately trained, retraining is unnecessary unless the user intends to alter the priors or assumes different noise characteristics.

The speed at which posterior samples are generated by all samplers used, including Normalizing Flows, is outlined in Table~\ref{table:efficiency}.
The runtime for the benchmark samplers is defined as the time it takes to complete their analyses. 
In the case of our model, this time is measured as the total time required to generate 10,000 samples.

Applying our model to the same test case of an MBHB signal, our model generates samples from the posterior at a rate approximately three orders of magnitude faster than our benchmark analyses using traditional inference techniques.
These traditional techniques, serving as a benchmark, were run on 90 CPU cores, showcasing the computational demands of conventional methods. 
This significant enhancement in computational efficiency underscores the potential of our approach in dealing with complex, high-dimensional data in gravitational wave data analysis and further emphasizes the advantages over traditional methods that require substantial computational resources.

\section{Conclusions and future prospects}\label{Sec:final}

In this study, we have showcased the first-of-its-kind application of machine learning techniques, specifically Normalizing Flows, to the realm of space-based gravitational wave detection, with a focus on the Taiji mission. 
This integration of advanced computational techniques with space-based observations marks a significant stride in the field. 
Our method offers a comprehensive statistical inference for MBHBs, setting a new benchmark in space-based gravitational wave astronomy. 
Notably, our approach remains resilient against challenges such as confusion noise and the intricacies of varying reference times, underscoring its reliability and potential for broader applications. 
The remarkable speed of our method, which outpaces traditional techniques by several orders of magnitude, establishes it as an invaluable tool for preprocessing in global fitting. 
This acceleration in data processing and analysis not only streamlines workflows but also broadens the horizons for gravitational wave data exploration. 
Together, these advancements herald a transformative era in space-based gravitational wave detection, fortifying the capabilities of missions like Taiji and edging us closer to deciphering the mysteries of the universe.

    % ...
Looking ahead, several promising avenues beckon our exploration. We aim to broaden our source considerations by extending to a more encompassing prior and refining our waveform, drawing inspiration from studies such as~\cite{pratten2023precision}.
This would provide a deeper understanding of the implications on inferred parameters.
A comparative analysis with works like~\cite{Cornish:2021smq} and~\cite{Weaving:2023fji}, focusing on multi-source scenarios (overlapping MBHBs), is also on our radar.
Furthermore, we are keen on testing our methodology on a diverse array of sources, encompassing EMRIs, verification GBs, and various stochastic GW backgrounds.

On the technical front, we recognize the imperative to factor in intricate and time-varying instrumental noise, data anomalies such as gaps and glitches, other TDI channels and generations, as well as more realistic satellite orbits, each of which adds layers of complexity to the whole problem.
Additionally, our current framework does not provide for evidence calculation, a gap we intend to bridge in subsequent research.

Our future endeavors will delve into understanding the ramifications of window length and devising strategies to infer the inspiral stage (approximately 2 weeks pre-merger) of MBHBs.
Such insights would be instrumental for alert systems and would further enrich global fitting processes.
We are also poised to explore methodologies to infer residuals from the global-fit approach.

Lastly, in light of the feedback from the reviewers, we acknowledge the importance of considering a more accurate prior range for spin parameters, specifically a physical prior of [-0.99, 0.99]. The current prior bound of [0, 0.99] was chosen as a compromise for handling complex data at this preliminary stage of estimation. However, we recognize that this simplification may bias the estimation of crucial parameters such as masses and coalescence time in real data scenarios. Going forward, we aim to improve our model to accommodate the complete spin range, which we anticipate will yield more accurate results. Moreover, this amendment will provide an opportunity to further investigate the degeneracy between spin and other parameters within the context of machine learning-based parameter estimation. We are grateful for the discerning observations from the anonymous reviewer which have guided this crucial insight.

By addressing these challenges and expanding our horizons, we aspire to not only refine the precision and robustness of our techniques but also pave the way for a holistic understanding of space-based gravitational wave data in the forthcoming era.



\Acknowledgements{
We are thankful to Zhoujian Cao for many helpful discussions.
This work was supported by the National Key Research and Development Program of China (Grant No. 2021YFC2203004 and Grant No. 2021YFC2201903).
H.W. is supported by the National Natural Science Foundation of China (NSFC) (Grant Nos. 12147103, 12247187) and the Fundamental Research Funds for the Central Universities. Minghui Du and Bo Liang collaborated extensively on the code debugging and results compilation process. Minghui Du took charge of preparing the Taiji datasets and developing the software used for training, testing, and inference. Minghui Du introduced the innovative concept of transforming reference times, enriching the methodological approach.
Bo Liang conducted an independent inference study to validate the reproducibility of our model. Bo Liang contributed significantly to code debugging and model fine-tuning, ensuring the robustness of the methodology.
He Wang assumed a leadership role, spearheading the overall project and coordinating the manuscript's writing. He Wang also played a pivotal role in verifying the credibility of the model's unbiased estimation and validating the multimodality of the posterior distribution.
Peng Xu made invaluable contributions by refining the narrative structure of the manuscript and enhancing its logical progression. Peng Xu's insightful suggestions greatly enriched the overall quality of the paper.
Ziren Luo and Yueliang Wu provided foundational contributions that set the tone for the manuscript. Their valuable insights and feedback played a crucial role in shaping the writing and submission process.
Yueliang Wu secured the major funding for this research endeavor and conceived the overarching research direction, providing the visionary framework within which the study unfolded.
All authors actively participated in the development of ideas, as well as the writing and rigorous reviewing of this manuscript. 
}%


\InterestConflict{The authors declare that they have no conflict of interest.}

\bibliographystyle{BibStyle}
\bibliography{main}

\begin{appendix}

\section{Model configuration and hyperparameters} \label{appendix:modelconfig}


In the proposed method, high-dimensional gravitational wave data are processed through an Embedding Neural Network (ENN), followed by a Neural Spline Flow (NSF). 
The details of both components are described below.


\textbf{Embedding Neural Network:}
The embedding neural network used in our study consists of 32 residual blocks, containing hidden layers that diminish in size, ranging from 1024 to 128 dimensions. 
The processed data is eventually transformed into a 128-dimensional vector. 

\textbf{Neural Spline Flow:} 
Following the ENN, the vector is further processed by the NSF, comprised of 30 flow steps with 1024-dimensional hidden layers, organized into 5 transformation blocks. 
The Exponential Linear Unit (ELU)~\cite{clevert2015fast} function is the activation function, and the `rq-coupling' function serves as the base transform type in the \lstinline|nflow| Python package~\cite{nflows}. 
The dropout probability is 0, and batch normalization is applied. The spline flow employs 8 bins for data discretization.

A summary of key hyperparameters for both the ENN and the NSF is provided in Table \ref{table:hyperparameters}.

\begin{table}[H]
    \centering
    \caption{Key hyperparameters of the networks.}
    \label{table:hyperparameters}
    \begin{tabular}{p{1cm}cc}
        \hline
        \hline
        \textbf{network} & \textbf{hyperparameter} & \textbf{value} \\
        \hline
        \multirow{3}{*}{\textbf{ENN}} & Residual Blocks Count & 32 \\
        & Hidden Layers Sizes & 1024, 512, 256, 128 \\
        & Final Vector Dimension & 128 \\
        \hline
        \multirow{8}{*}{\textbf{NSF}} & Flow Steps & 30 \\
        & Hidden Layer Size & 1024 \\
        & Transform Blocks & 5 \\
        & Activation & elu \\
        & Dropout & 0.0 \\
        & Batch Norm & True \\
        & Bins & 8 \\
        & Base Transform & rq-coupling \\
        \hline
        \hline
    \end{tabular}
\end{table}

\section{Comparison of approximate posterior distributions} \label{appendix:full_corner}

\begin{figure*}[t] 
  \centering
  \includegraphics[width=1\textwidth]{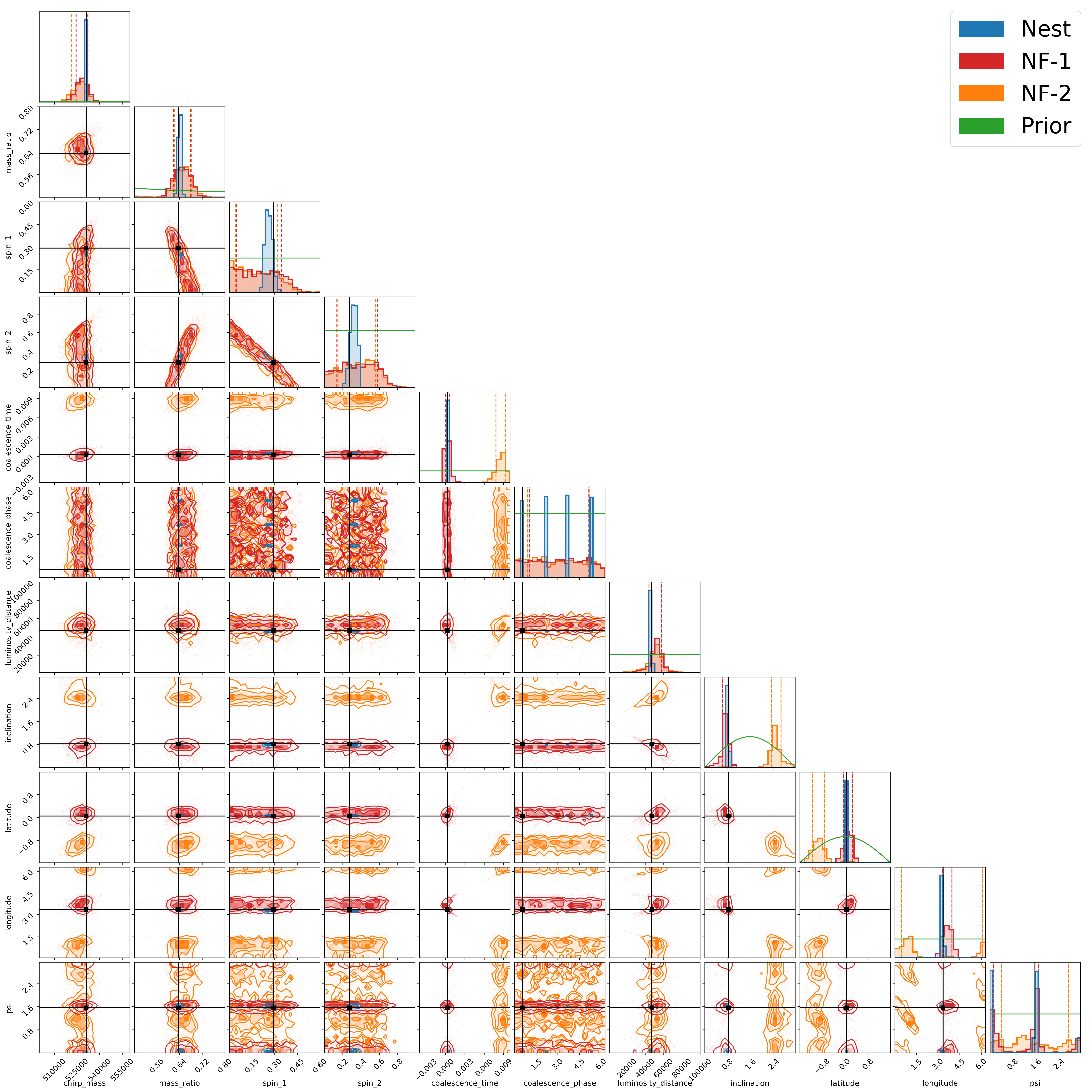}
  \caption{
    This corner plot displays the one and two-dimensional marginalized posterior distributions of the GW parameters for a selected example.
    The red and orange contours represent the two-dimensional joint posteriors from our NF model, labeled as NF-1 (dominant or best-fit) and NF-2 (subdominant), respectively. The blue contour illustrates the corresponding posteriors derived from the Nested Sampling (Nest) method.
    The contour boundaries for each method define the $1\sigma$ level.
    The one-dimensional histograms of the posterior distribution for each parameter from both methods are plotted on the diagonal, with the prior distribution also depicted in green.
    Black vertical and horizontal lines mark the true parameter values of the simulated signal.
    This visualization facilitates a comprehensive comparison and validation of the NF model against the conventional stochastic sampling method, revealing the observed multimodality in the NF posteriors.
  }
  \label{fig:corner}
\end{figure*}

In this Appendix, we present a detailed comparison between the full 11-dimensional posteriors that we have analyzed. 
Specifically, we focus on the 90\% marginal distributions, highlighting the similarities and differences between our normalizing flow (NF) model and the traditional Nested Sampling method. 
Figure~\ref{fig:corner} illustrates this comparison, where the largely overlapping distribution between the Nested Sampler (in blue) and the NF model (in red and orange) is apparent. 
Additionally, the prior distribution (in green) is plotted for a clear and direct comparative perspective.

\begin{figure*}[t] 
  \centering
  \includegraphics[width=0.8\textwidth]{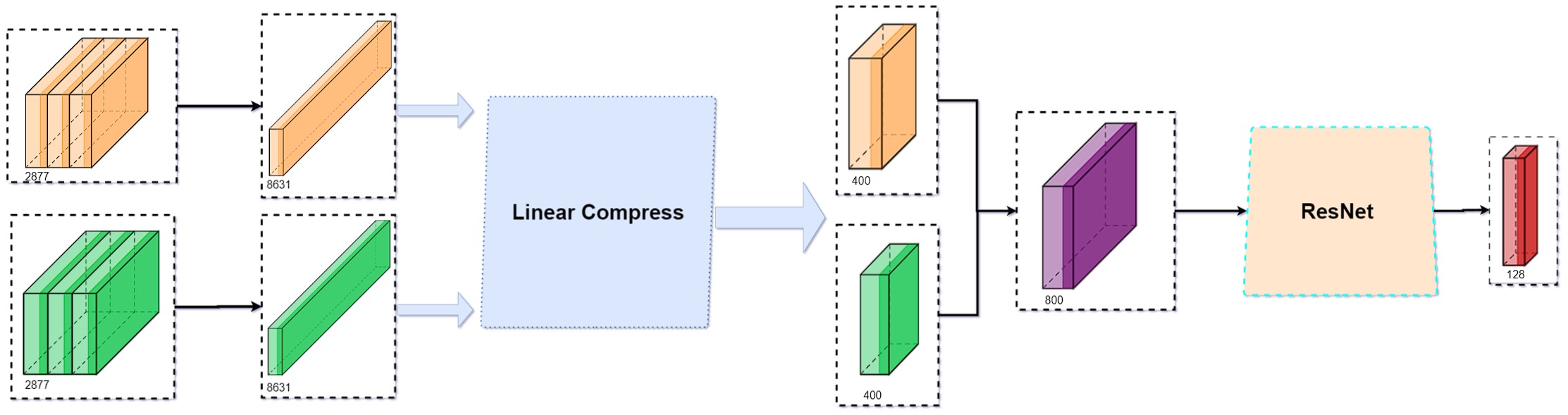}
  \caption{A schematic representation of the framework for compressing high-dimensional gravitational wave data.
  In this diagram, ``Yellow data block" and ``Green data block" respectively represent the data from detectors A and E. Both detectors house two-dimensional information as (3, 2877), corresponding to the sequences of the real and imaginary parts, and Amplitude Spectral Density (ASD), each with a length of 2877.
  The sequences are concatenated to reach a length of 8631 and undergo linear compression, reducing each to 400 dimensions. The compressed data are concatenated into an 800-dimensional representation and processed through a ResNet for feature extraction, compressing the entire gravitational wave data from (2, 3, 2877) to 128 dimensions.
  }
\end{figure*}

The inclusion of both dominant (NF-1) and subdominant (NF-2) contours from our NF model in Figure~\ref{fig:corner} allows for an intricate examination of the model's performance, emphasizing its capabilities and alignment with Nested Sampling.
The integration of multiple modalities in the NF posteriors further contributes to the robustness of our analysis and offers valuable insights into the complexities of the gravitational wave data.

\end{appendix}

\end{multicols}

\end{document}